
\input phyzzx.tex
\PHYSREV
\pubnum={IFUM 450/FT}
\date={}

\def\la{\Lambda}
\def\ala{|\Lambda|}
\def\ep{\epsilon}
\def\ff{\phi}

\def\fat{{1 \over {(16 \pi G)}}}

\def\lar{\lambda (r)}
\def\epr{\epsilon (r)}
\def\mur{\mu (r)}
\def\nur{\nu (r)}
\def\ffr{\phi (r)}
\def\vfr{\varphi (r)}

\def\pla{{\lambda}^{\prime}}
\def\ppla{{\lambda}^{ \prime \prime}}
\def\pg{g^{ \prime}}
\def\pmu{{\mu}^{ \prime}}
\def\pep{{\epsilon}^{ \prime}}

\def\pnu{{\nu}^{ \prime}}
\def\pfi{{\phi}^{ \prime}}
\def\ppfi{{\phi}^{ \prime \prime}}
\def\pf{{\varphi}^{ \prime}}
\def\ppf{{\varphi}^{ \prime \prime}}
\def\ppmu{{\mu}^{\prime \prime}}

\def\gtad{{\tilde g}_{a d}}
\def\gtbc{{\tilde g}_{b c}}
\def\gtac{{\tilde g}_{a c}}
\def\gtbd{{\tilde g}_{b d}}

\def\fab{{\phi}_{a b}}
\def\fad{{\phi}_{a d}}
\def\fbc{{\phi}_{b c}}
\def\fac{{\phi}_{a c}}
\def\fbd{{\phi}_{b d}}

\def\gtab{{\tilde g}_{a b}}

\def\gtmn{{\tilde g}_{\mu \nu}}
\def\gmn{g_{\mu \nu}}
\def\goo{g_{00}}
\def\gii{g_{11}}

\def\rtil{{\tilde R}_{abcd}}
\def\ftil{{\tilde \Phi}_{abcd}}
\def\gtil{{\tilde G}_{abcd}}

\def\Rmn{R_{\mu \nu}}
\def\Ri{R_{abcd}R^{abcd}}

\def\inte{\int d^{4}x {\sqrt {-g}}}

\def\intev{\int_{V} d^{4}x {\sqrt {-g}}}

\def\xo{\tilde x}

\def\intt{\int d^{2}{\tilde x} {\sqrt {-{\tilde g}}}}
\def\intt{\int d^{2}{\tilde x} {\sqrt {-{\tilde g}}}}

\def\dmu{\nabla_{\mu} \phi}
\def\dnu{\nabla_{\nu} \phi}
\def\dqu{(\nabla \phi)^2}

\def\dqvt{({\tilde {\nabla}} \varphi)^2}
\def\dqut{({\tilde {\nabla}} \phi)^2}

\def\nabta{{\tilde {\nabla}}_{a}}
\def\nabtb{{\tilde {\nabla}}_{b}}
\def\nabt{\tilde {\nabla}}

\def\dll{\nabla^{2} }

\def\dla{\nabla^{2} \phi}
\def\mev{e^{- 2 \varphi}}
\def\mef{e^{- 2 \phi}}
\def\medf{e^{- 2 \phi}}
\def\edv{e^{2 \varphi}}
\def\edf{e^{2 \phi}}
\def\edla{e^{2 l}}
\def\edl{e^{2 \lambda}}

\def\edn{e^{2 \nu}}

\def\scw{Schwarzschild}
\def\scwd{Schwarzschild-de Sitter}
\def\scwa{Schwarzschild-anti de Sitter}
\def\des{de Sitter}
\def\astr{\alpha^{\prime}_{string}}

\titlepage
\title{Black Holes and Cosmological Constant in Bosonic String Theory: Some
 Remarks}

\author{F. Belgiorno\foot{Also Sezione I.N.F.N. di Milano,
 20133 Milano, Italy. E-mail: belgiorno@vaxmi.mi.infn.it}
}
\address{Dipartimento di Fisica, Universit\`a di Milano, 20133 Milano, Italy}

\andauthor{A.S. Cattaneo\foot{Also Sezione I.N.F.N. di Milano,
 20133 Milano, Italy. E-mail: cattaneo@vaxmi.mi.infn.it}}
\address{Dipartimento di Fisica, Universit\`a di Milano, 20133 Milano, Italy}

\abstract{In this work we attempt to give some suggestion about the black
hole solutions coming from the bosonic string theory with cosmological
constant  in four dimensions.\par
\noindent In our study we are first interested in the effects on the metric of
 a tree level cosmological constant; then we make a
perturbative calculation to be compared with the general relativistic results.}

\endpage

\chapter{Introduction}

There are essentially two questions we would like to find an answer to:\par
i)What about nontrivial dilaton physics in black hole solutions of the
 bosonic string theory with cosmological constant?\par
ii)What about the effects of a cosmological constant on the black hole
temperature?\par
There are various possibilities to implement this attempt: the off-critical
string theory gives  $\Lambda \propto D-26 $, so we can introduce it directly
to the tree level; another possibility is to think $\Lambda$ to the first order
in
$\alpha^{\prime}_{string}$.\par
\noindent
Our most interesting results are:\par
\noindent
\item {1)}{ introducing a non perturbative cosmological constant $\Lambda$
gives rise to
a nontrivial dilaton $\phi$: the solution $\phi = const$ (the only one if
we want black hole solutions) is not possible.\par
\noindent This means that if we are going to believe in the off-critical string
physics we find in the Einstein framework the following
results:}\par
$$\eqalign{
\phi &\neq const \cr
g_{00} &\neq -{g_{11}}^{-1} \cr}
$$
\item {2)}If we treat perturbatively $\Lambda$ and to the first order
in string
theory, as such as the square of the Riemann tensor, we still find nontrivial
corrections to the geometry and the black hole temperature; this
non-triviality is a cosmological constant effect.\par
\noindent For completeness a parallel study with the general relativistic
 Schwarzschild-De Sitter solution is made.

\chapter{Non-perturbative Analysis}

The action we are interested in is the effective bosonic string
action in $4D$, with $\la \neq 0$ (the case with $\la = 0$ is already treated
 in Callan-Myers-Perry
\Ref\CMP{C.G. Callan, R.C. Myers and M.J. Perry\journal Nucl. Phys.
&B311(89)673.}[CMP]):
$$
S= \fat \inte \mef (R + 4 \dqu - 2 \la + {k \over 2} \Ri),
\eqn\azio
$$
\noindent The last term in this action is the well known first perturbative
correction coming from the string beta function equations.\par
\noindent The perturbative expansion parameter k is related to $\astr$ by
the relation
$$
k = {\astr \over 2}
\eqn\kkk
$$
Let us make a Weyl transformation so as to go into the so-called Einstein
framework:
$$
\gmn \rightarrow \edf \gmn
\eqn\conf
$$
and let us assume at first that $\la$ is not perturbative.\par
\noindent
The transformed action to the zero order is
$$
S_{E}= \fat \inte (R - 2 \dqu - 2 \la \edf)
\eqn\taz
$$
and gives rise to the following equations:
$$
\Rmn - 2 \dmu \dnu - \gmn \la \edf = 0
\eqn\nonpt
$$
$$
\dla - \la \edf = 0
\eqn\non
$$
We are looking for sphero-symmetric statical solutions, so we put\par
$$
\eqalign{
ds^{2} &=  - \edl dt^{2} + \edn dr^{2} + r^{2} d {\Omega }^{2}\cr
\lambda &= \lar ; \nu = \nur ;   \phi = \ffr \cr}
$$
Substituting in \nonpt\ and in \non\ and denoting with a prime the derivative
with respect to r, we find:
$$
\eqalign{
 \pnu \pla - \ppla - {\pla}^2 - {2 \over r} \pla &= \la \edf \edn ,\cr
 \pnu \pla - \ppla - {\pla}^2 + {2 \over r} \pnu - 2 {\pfi}^2
&= \la \edf \edn ,\cr
r( \pnu - \pla) - 1 + \edn &=
r^2 \la \edf \edn ,\cr
\ppfi +( \pla - \pnu + {2 \over r}) \pfi &= \la \edf \edn , \cr}
\eqn\fond
$$
Subtracting the first equation in \fond\  from the second one  we have
$$
(\pla + \pnu) - r {\pfi}^2 = 0.
\eqn\sot
$$
{}From \sot\  one can realize that the choice
$$
\lambda = -\nu \Leftrightarrow \goo = - {\gii}^{-1}
\eqn\gogi
$$
is incompatible with \non :
$$
\lambda = -\nu \Leftrightarrow \pfi = 0 \Rightarrow \la = 0 \hbox{ in \non}
\eqn\nogo
$$
So, taking a non-perturbative $\la$, we conclude that:\par
\item{a)}
the solution $\ff = const$ is not allowed.
\item{b)}
$\goo \neq - {\gii}^{-1} $.
This non-triviality of the solutions is a cosmological constant effect:
without a non-perturbative $\la$ the only one black hole solution,
as we shall show in the appendix, is
a constant dilaton in a \scw\  metric.
\noindent We are not able to find an analytical solution, so we cannot say
 anything about the effective existence of a black hole solution for a tree
 level cosmological constant; the relevant point is the obvious absence of
matching with the General Relativity.

\chapter{Perturbative Approach}

Lacking any analytical black-hole solution for non-perturbative $\la$,
 let us put
$$
\la \equiv \alpha k.
\eqn\cost
$$
This corresponds to introduce the cosmological constant to the first order
in the string Lagrangian.\par
\noindent To the tree level, as in CMP, the action is
$$
S_{E}= \fat \inte (R - 2 \dqu)
\eqn\se
$$
and the corresponding equations
$$
\eqalign{
\Rmn - 2 \dmu \dnu &= 0 \cr
\dla &=0 \cr }
\eqn\moto
$$
allow as the only sphero-symmetric static solution under the reasonable
 boundary condition
$$
\ffr \rightarrow {\phi}_{0} + {A \over r} + O({1 \over r^2}),
\eqn\bc
$$
the CMP solution:
$$
\biggl\{
\eqalign{&{\rm {\scw\  metric}}\cr
&\ff= {\phi}_{0} = const\cr}
\eqn\sot
$$
\noindent We can put without any loss of generality $\phi_{0}=0$.
The perturbative correction is
$$
S^{(1)}= \fat k  \inte (- 2 \alpha \edf + {1 \over 2} \medf [\Ri+ \ldots] )
\eqn\pt
$$
\noindent The dots in \pt\  mean further terms involving higher derivative
terms of the dilaton field; with a field redefinition of $\ff$ to the first
order in k it is possible to eliminate these terms; in the appendix we shall
give further technical details on these stuff.\par
In the sequel we will again indicate the redefined dilaton with $\ff$.\par
The equations to the first order are
$$
\eqalign{
\Rmn^{(1)}- \alpha \gmn^{(0)} + [\Rmn] - {1 \over 4}
\gmn^{(0)} [R]^{2} &=0, \cr
\dla^{(1)} - \alpha -{1 \over 4} [R]^{2} &= 0 \cr }
\eqn\fst
$$
where
$$\eqalign{
[\Rmn] &\equiv R_{\mu abc} {R_{\nu}}^{abc},\cr
[R]^{2} &\equiv \Ri, \cr}
$$
we put
$$\eqalign{
\lar &= l(r) + k \mur \cr
\nur &= -l(r) + k \epr \cr
\edla &= g(r) = 1- {{2 m} \over r}\cr
\ffr &=	\phi_{0} + k \vfr \equiv \phi_{0} + k \phi^{(1)} \cr}
$$
\noindent substituting in \fst\  the equations become
$$
{{r - 2 m} \over r} \ppf + 2 {{r - m} \over r^{2}} \pf - 12 {{m^2} \over
 r^{6}} - \alpha =0
\eqn\dilf
$$
\noindent and
$$
\eqalign{
- m \pep + (2 r - m) \pmu + (r - 2 m) r \ppmu + \alpha r^{2} &=0 \cr
(3 m - 2 r) \pep + 3 m \pmu + r (r - 2 m) \ppmu + \alpha r^{2} &=0 \cr
2 \ep + (r - 2 m) \pep - (r - 2 m) \pmu -\alpha r^{2} &=0 \cr }
\eqn\first
$$
\noindent
As physical boundary conditions we require the perturbations to be
regular on the horizon $r_{H} = 2 m $; the cosmological term prevents us
 to require the asymptotic flatness.\par
\noindent For the dilaton we find:\par
$$\eqalign{
\vfr = &{\varphi}_{CMP}(r) + {4 \over 3} m^2 \alpha log(r) + \cr
       & {2 \over 3} m \alpha r + {1 \over 6} \alpha r^2 + const, \cr}
\eqn\soldil
$$
where
$$
\varphi_{CMP}(r) \equiv -{{2 m} \over {3 r^3}}-{1 \over {2 r^2}}
-{1 \over {2 m r}},
\eqn\cmpsol
$$
\noindent is the CMP solution.\par
\noindent

The dilaton solution is regular on the horizon, but it is not asymptotically
 flat.\par
\noindent
The three equations for $\mur, \epr$ are functionally related by means of
 the Bianchi identity\rlap;\Ref\cmpf{C.G. Callan,
E.J. Martinec,M.J. Perry and D. Friedan\journal
Nucl. Phys. &B262(85)593.}
subtracting the first equation in \first\  from the second one it follows\par
$$
(r - 2 m)(\pep + \pmu)=0,
\eqn\epmu
$$
so
$$
\mu = - \ep,
\eqn\go
$$
and the third equation in \first\  gives
$$
\epr = {\alpha \over 6}(r^2 + 2 m r + 4 m^2).
\eqn\solep
$$
The divergence of the perturbations for $r \rightarrow \infty$ shows that
 exists a validity limit of the perturbative expansion.\par
Finally the metric components are\par
$$\eqalign{
\goo &= - \edl = - g(r) \cdot (1- {\la \over 3}(r^2 + 2 m r + 4 m^2))+
O({\la}^{2}), \cr
\gii &=  \edn = - {1 \over \goo} =
{1 \over g(r)} \cdot (1+ {\la \over 3}(r^2 + 2 m r + 4 m^2))+
O({\la}^{2}), \cr}
\eqn\metcom
$$
The structure of the unperturbed space-time
is consistently modified by the
 perturbative cosmological term; besides, if in the formula
$$
\goo = - g(r) \cdot (1+2 k \mur)
$$
\noindent we can think to extend the validity of the results in the region
where\foot{some reason to make this extension could come from the analogy
 with the \scw\  case, where a calculation perturbative in $\la$
 lets guess the existence of a cosmological horizon; see also sec. 5}
$$
|2 k \mur|\sim 1,
$$
then for $\la > 0$ there exists the possibility
 to have a second horizon: the new zeroes of $\goo$ given by
$$
r^2 + 2 m r + 4 m^2 - {3 \over \la} = 0
\eqn\zer
$$
which means
$$
\eqalign{
r_{+} &= -m + {\sqrt {3}} {\sqrt {{1 \over \la} - m^2}} \cr
r_{-} &= -m - {\sqrt {3}} {\sqrt {{1 \over \la} - m^2}} \cr}
\eqn\rad
$$
The reality condition imposes
$$
\Delta \equiv {1 \over \la} - m^2 \geq 0 \Leftrightarrow
m \leq {1 \over {\sqrt \la} }
\eqn\re
$$
with the inequality needed for to be $r_{+} > r_{-}$.
We actually have a new horizon only for $r_{+} > 0$, that is
$$
m \leq {{\sqrt 3} \over 2} {1 \over {\sqrt \la}}.
\eqn\hor
$$
\noindent Making a series expansion of $r_{+}$ in $\la$ for $\la \rightarrow 0$
$$
r_{+} = - m + {{\sqrt 3} \over {\sqrt \la}} - {{\sqrt 3} \over 2} {\sqrt \la}
m^2 + O(\la),
\eqn\rpiu
$$
and taking $\la$ small enough, it results $r_{+} > r_{H}$, so one can identify
$r_{+}$ as a cosmological horizon like the \des\  horizon in the \scwd
\  solutions\rlap.\Ref\GH{G.W. Gibbons and S.H. Hawking\journal Phys. Rev.
&D10(1977)2738.}
\par
\noindent
Then, if \hor\  is satisfied, we find, as in the \scwd\  case,
two horizons;
it is reasonable to think that the absence of existence conditions for
 the black-hole horizon $r_{H}$ is a pure artifice of the perturbative
approximation: in the \scwd\  solution, as it is known, if
$m > {1 \over {3 \sqrt \la}}$, there doesn't exist any horizon and there is
a naked singularity\rlap;\foot{We stress anyhow that the appearance of a naked
singularity,
taking $\la \sim 3 \cdot 10^{-52} meters^{-2}$ (the experimental upper-bound
for $|\la|$), would occur for
$$
m_{critical} = {1 \over {3 \sqrt \la}} \sim 10^{22} {\rm solar\  masses}
\eqn\macri
$$
 a value very near the universe mass; so from a physical point of view we
could avoid to be worried about it.}
something similar could happen also in our case.\par

The case where $\la < 0$ is also interesting; we then have only the horizon
 $r_{H}$; it is natural to make a parallel between our string solutions and the
\scwa\  exact solutions of the General Relativity, in which the only horizon
occurs for
$$
r^{SAD}_{H}= ({{3 m} \over \ala})^{1 \over 3}
((1 + {\sqrt {1+{1 \over {9 m^2 \ala}}}})^{1 \over 3}+
(1 - {\sqrt {1+{1 \over {9 m^2 \ala}}}})^{1 \over 3})
\eqn\scar
$$

\chapter{Semiclassical Aspects}

The appearance of non-trivial perturbations affects the black-hole
temperature by means of the cosmological contributions.\par
\noindent We remember that for $\la = 0$ (cfr. CMP) in 4D there is no
new contribution to the black-hole temperature beyond the Hawking one.
There are two cases:\par
\noindent 1)$$\la > 0$$\par
with two subcases:\par
\noindent a)$$m < {{\sqrt 3} \over 2} {1 \over {\sqrt \la}} ;\eqn\aa$$\par
\noindent b)$$m > {{\sqrt 3} \over 2} {1 \over {\sqrt \la}}\eqn\bb$$\par
\noindent
In the case a), \aa,
the surface gravity K is to be calculated for the black-hole horizon and
 for the \des\  horizon; we find respectively
$$
K_{H}={1 \over {4 m}}|1-4 m^2 \la|+O({\la}^{2}),
\eqn\kh
$$
and
$$
K_{C}={{\sqrt {\la}} \over {\sqrt 3}}|1-
{2  \over {\sqrt 3}}{\sqrt {\la}} m|+O({\la}^{3 \over 2});
\eqn\kc
$$
then
$$
T_{H}={1 \over {8 \pi m}}|1-4 m^2 \la|+O({\la}^{2})
\eqn\th
$$
and
$$
T_{C}={{\sqrt {\la}} \over {2 \pi \sqrt 3}}|1-
{2  \over {\sqrt 3}}{\sqrt {\la}} m|+O({\la}^{3 \over 2}).
\eqn\tc
$$
\noindent We can compare these temperatures with those obtained from the
\scwd\  [SD] case in the limit $\la \rightarrow 0$:
$$
T_{H}^{SD}={1 \over {8 \pi m}}|1-{16 \over 3} m^2 \la|+O({\la}^{2}),
\eqn\ths
$$
$$
T_{C}^{SD}={{\sqrt {\la}} \over {2 \pi \sqrt 3}}|1-
{2  \over {\sqrt 3}}{\sqrt {\la}} m|+O({\la}^{3 \over 2}).
\eqn\tcs
$$
The contributions are very similar.

{}From a physical point of view\refmark\GH the temperature actually measured
by one
 observer is given by a mixture of the black-hole and the cosmological
radiations\rlap.\foot{
The cosmological temperature, that one observer should see coming from
any direction in the universe, would have the
 following upper bound:
$$
T_{C} \sim {{\sqrt {\la}} \over {2 \pi \sqrt 3}} {{h c} \over {2 \pi
k_{B}}} < 2.3 \cdot 10^{-29} \ {}^{\circ}\!K
$$ }

\noindent In the case b), \bb, there is only the black-hole horizon,
with temperature given by \th.

In the \scwd\  case, we know that both the temperatures decrease
\GH\ when m increases, so they cannot diverge with m as it seems by looking
at our approximates formulas; anyway, the existence condition of the event
horizons $m < {1 \over {3 \sqrt \la}}$ is such that
$$\eqalign{
{16 \over 3} \la m^2 &< 1 \cr
{2 \over {\sqrt 3}}{\sqrt \la} m &<1 \cr}
\eqn\bound
$$
so \ths\  and \tcs\  cannot diverge with m and they
are actually decreasing functions of m.
\noindent We can see a similar mechanism
provided by an horizon existence condition
prevents that $T_{H}$ and
$T_{C}$ diverge with m also in our string framework in the case a).
\par
\noindent In the case b), there is a divergence probably due to the
perturbative approximation.

\noindent 2)$$\la < 0$$\par

\noindent It results:
$$
K_{H}={1 \over {4 m}}(1+4 m^2 \ala)+O({\la}^{2}),
\eqn\kha
$$
and
$$
T_{H}={1 \over {8 \pi m}}(1+4 m^2 \ala)+O({\la}^{2}),
\eqn\tha
$$
whereas the \scwa\  solution gives
$$
T_{H}^{SAD}={1 \over {8 \pi m}}(1+{16 \over 3} m^2 \ala)+O({\la}^{2})
\eqn\thsa
$$
The formulas \tha\  and \thsa\  show a pathological behaviour
for $m \rightarrow \infty$: we cannot appeal to any principle to avoid
the temperature divergence in the limit $m \rightarrow \infty$.\par
\noindent We can say that this behaviour physically begins to occur for
very large masses (order of the universe mass).

\chapter{On General Relativity Analogies}

We want to make the same perturbative approach in $\la$
for the General Relativity although the exact solution are well known:
we could learn something about our solutions.\par
\noindent Using exactly the same definitions as in section 3, we write the
equations at the first order in k:
$$
\eqalign{
{\pg \over {2g}}(\pep - \pmu)-\ppmu -
{\pg \over {g}}\pmu -{2 \over r}\pmu -{\alpha \over g} &=0 \cr
{\pg \over {2g}}(\pep - \pmu)-\ppmu -
{\pg \over {g}}\pmu +{2 \over r}\pep -{\alpha \over g} &=0 \cr
r(\pep - \pmu)+2 {\epsilon \over g}-r^{2}{\alpha \over g} &=0 \cr}
\eqn\dept
$$
\noindent Subtracting the first equation in \dept\  from the second one
we get
$$
{2 \over r}(\pep + \pmu)=0,
\eqn\desgo
$$
so again
$$
\mu = - \ep,
$$
and the third equation in \dept\  gives
$$
2rg\pep + 2 \ep - r^{2} \alpha=0,
\eqn\epsi
$$
which is the same equation we obtained in sec. 3 : so we conclude that a
perturbative approach to the \scwd\  physics leads to the same results for
the metric as from the \se\  action.\par
\noindent Actually what is ``new" in the string action in comparison with
the General Relativity is the coupling with the dilaton field; it occurs in eq.
I) of sec. 2 by means of ${\pfi}^{2}$ and $\la \edf$.\par
If we put
$$
\phi=\phi_{0} + k \varphi(r)
\eqn\costdil
$$
\noindent (and this is consistent only for a perturbative $\la$) then
$$\eqalign{
\la \edf &= \alpha k e^{2 \phi_{0}} + O(k^2) = \alpha {\tilde k}  +O(k^2),\cr
{\pfi}^{2} &= O(k^2);\cr}
\eqn\cou
$$
that is, apart from a renormalization of the coupling constant (but we
can choose $\ff_{0}=0$, as we did in sec. 3) the
dilaton decouples from the equations for the metric perturbations to order
k.\par
\noindent So the only possibility to appreciate the presence of the
dilaton field seems to be confined to the case of a non trivial $\phi$
to the tree level;
this is possible for a non perturbative cosmological constant.\par
\noindent Nevertheless the matching with the analogous general relativistic
calculations is not yet automatic, because of the $\Ri$ stringy contribution;
but what actually
happens is that on-shell
$$
[\Rmn] = {1 \over 4}
\gmn^{(0)} [R]^{2},
\eqn\onshell
$$
so it disappears from the equations for $\ep$ and $\mu$.\par
We can in this way understand the equivalence of the results obtained from the
General Relativity and from the string action in the Einstein framework.

\chapter{Quantum Aspects and Conclusions}

Not being possible to quantize the four dimensional string action, taken in the
 Einstein framework, we can think to make a ``dimensional reduction" of it :
we choose a sphero-symmetric ansatz for the 4D-metric
$$
ds^{2}=\gtab (\xo^{0},\xo^{1}) d\xo^{a}d\xo^{b} +
e^{- 2 \varphi (\xo^{0},\xo^{1})} d{\Omega}^{2} \hskip .4truecm a,b=0,1
\eqn\rid
$$
and
$$
\phi = \phi(\xo^{0},\xo^{1})
\eqn\dd
$$
The zero order action in k \se\  becomes
$$
S_{E} \rightarrow S_{2}= \intt \mev ( {\tilde R} + 2 \dqvt + 2 \edv
- 2 \dqut )
\eqn\sdil
$$
\noindent This reduced action is a variant of dilaton gravity
\Ref\LN{O. Lechtenfeld and C. Nappi\journal Phys. Lett. &B288(92)72}
 minimally coupled
 with a scalar field $\phi$ which is the real 4D-dilaton field; there
still exists
 a black-hole solution of classical equations with cosmological perturbation
which is equal to the four-dimensional one.\par

The idea is to quantize this action, following the common ansatz that the 4D
 angular degrees of freedom qualitatively don't modify the bidimensional
picture; but unfortunately we cannot implement this program because with \sdil
\  we are not at the critical conformal point: the dilaton gravity action
conformally invariant is the Callan, Giddings, Harvey and  Strominger (CGHS)
\Ref\CGHS{C.G. Callan, S.B. Giddings, J.A. Harvey and A. Strominger\journal
Phys. Rev. &D45(92)R1005.}
action:
$$
S_{CGHS}= \intt \mev ( {\tilde R} + 4 \dqvt + 4 {\lambda}^{2})
\eqn\cghs
$$
The very crucial differences are sketched by the following rules
$$
\eqalign{
 CGHS \longrightarrow &\sdil \cr
4 \dqvt \longrightarrow &2 \dqvt \hbox{ in \sdil}\cr
4 {\lambda}^{2} \longrightarrow &2 \edv \hbox{ in \sdil}\cr}
$$
\REF\za{M.T. Grisaru, A. Lerda, S. Penati and D. Zanon,
\journal Nucl. Phys. &B342(90)564.}
Besides, we have another scalar $\phi$ making more intricate the study.\par
\noindent However, the major problem to be faced if one wants to carry on this
 program is the non Toda-like form of our action, and as a consequence one
cannot use the standard conformal field theory methods employed in ref.
\za\ for the quantization of \sdil.\par

\vskip 0.8truecm
\REF\HH{J.H. Horne and G.T. Horowitz, {\sl Cosmic Censorship and the Dilaton},
hep-th/9307177, to appear in {\it Phys. Rev.} {\bf D}, Rapid Communications.}
In this work we have taken into account the physical contributions of the
cosmological constant\foot{After this work was completed, we became aware
of Ref. \HH, in which an explicit solution to string equations of motion
with a tree level cosmological constant, albeit in the extremally charged case,
is presented and the problem of finding an exact solution in the uncharged
case is discussed.}
in the framework of the bosonic string theory in four
dimensions; particularly, we have looked for statical sphero-symmetric
black hole solutions and we have carried out a parallel study with
the General Relativity [GR] to conclude that:\par
i) a tree level $\la$ gives no matching with [GR]\par
ii) a first order $\la$ gives the same results as a first order in $\la$
 [GR].\par
A semiclassical treatment of the black hole evaporation in the
limit $\la \rightarrow 0$ has shown that there is a cosmological contribution
 to the black hole temperature and that in the case of $\la < 0$ the
temperature has a pathological behaviour for large masses.\par

\ack
We thank M. Martellini for fruitful discussions and suggestions.

\endpage

\appendix

I) Black hole solutions in dilaton-gravity theories\par

\noindent We show that the only static sphero-symmetric
solution of the equations $A),B)$ of section 3 under suitable boundary
conditions
 compatible with a black-hole structure of the space-time is $\ff=const$;
we want to stress that the demonstration is quite the same as for the
scalar-tensor theories of gravity like Brans-Dicke theories \rlap.\ref{
S.W. Hawking\journal Commun. math. Phys. &25(1972)167}
We have the equation
$$
\dla=0
$$
under the boundary condition
$$
\ffr \rightarrow {\phi}_{0} + {A \over r} + O({1 \over r^2})
$$
\noindent Let us define
$$
\psi \equiv \ff-\ff_{0}
$$
\noindent obviously
$$
\dll \psi=0
$$
\noindent we have
$$
\eqalign{
\intev \psi \dll \psi &= 0 = \intev \nabla_{a} (\psi \nabla^{a} \psi )-\intev
(\nabla \psi)^{2} \cr
& = \int_{\Sigma_{\alpha}} \psi \nabla_{\alpha}
\psi d\Sigma_{\alpha} -\intev (\nabla \psi)^{2} \cr}
$$
\noindent where $\Sigma_{\alpha}$ stays for the hypersurfaces which bound the
volume V defined as follows:
$$\eqalign{
S =&{\rm spacelike\  hypersurface} \cr
S^{\prime} =&{\rm spacelike\  hypersurface\  translated\  along\
 the\  timelike\  killing}\cr
& {\rm vector\
of\  the\  exterior\  geometry} \cr
T =&{\rm timelike\  hypersurface\  at\  infinity} \cr
H =&{\rm event\  horizon} \cr}
$$
\noindent The surface integral over H is zero because $\nabla_{a} \psi$ has no
components along the symmetry directions and the surface element on H is
entirely in a Killing direction; the surface integrals over S and $S^{\prime}$
 cancel each other because of the time symmetry; the surface integral over T is
zero too because of the boundary condition on $\phi$. So we have
$$
\intev (\nabla \psi)^{2}=0
$$
\noindent Because the gradient of $\psi$ can be only spacelike or zero the
thesis follows.

II) On Weyl transformations\par

\noindent Under a conformal transformation
$$
\gmn = \edf \gtmn
$$
we have
$$
R_{abcd}=\edf ({\tilde R}_{abcd}+{\tilde \Phi}_{abcd}+{\tilde G}_{abcd}
\dqut)
$$
where
$$\eqalign{
\ftil &=\gtad \fbc + \gtbc \fad - \gtac \fbd -\gtbd \fac \cr
\gtil &=\gtad \gtbc - \gtac \gtbd \cr
\fab &=\nabta \nabtb \ff - (\nabta \ff)(\nabtb \ff) \cr}
$$
so
$$
\Ri=\mef (\rtil {\tilde R}^{abcd}+ f(\nabt \phi))
$$
\noindent
and f indicates a function of the dilaton derivatives; to eliminate this f we
make the following transformation:
$$
\ff \rightarrow \ff + k \cdot F(\phi)
$$
and choose F such that
$$
-2 \dqut + {k \over 2} \mef f(\nabt \phi) \rightarrow - \dqut + O(k^{2})
$$

\endpage\refout

\bye